\documentclass[12pt]{article}
\usepackage{epsfig}
\usepackage{hyperref}
\usepackage{amssymb}
\usepackage{colordvi}

\begin{document}

\begin{center}

{\Large\bf Comment on ``Revisiting the Wu-Yang approach to magnetic charge''}
\vspace*{0.4cm}

{\large
M.~G\"ockeler$^1$ and T. Sch\"ucker$^2$}

\vspace*{0.4cm}

{\sl
$^1$ Institut f\"ur Theoretische Physik, Universit\"at
Regensburg,  93040 Regensburg, Germany \\
$^2$ Aix Marseille Univ, Universit\'e de Toulon, CNRS, CPT, Marseille, France
}
\end{center}

\begin{abstract}
In two recent papers Gonuguntla and Singleton claim that the Wu-Yang
fiber bundle approach does not lead to to a consistent model for magnetic
charge. We point out that this claim is false.
\end{abstract}

In the paper~\cite{Gonuguntla:2023cik} Gonuguntla and Singleton revisit the 
Wu-Yang fiber bundle approach~\cite{Wu:1975es} to magnetic charge and come
to the conclusion that the Wu-Yang construction does not represent a true
magnetic charge, because it leads to additional contributions to the
magnetic field besides the monopole part. A similar statement is made
in another paper~\cite{Gonuguntla:2023uxi} devoted to the construction
of magnetic monopoles.

This claim is false. The essential error is Eq.(2) in
Ref.~\cite{Gonuguntla:2023cik} and the statement made in connection with
this equation that ``One can write the Wu-Yang fiber bundle potential
in a single expression''. This contradicts what Wu and Yang say in the
fourth paragraph of Section I of Ref.~\cite{Wu:1975es}, namely,
``that in general the phase factor (and indeed the vector potental $A_\mu$)
can only be properly defined in each of many overlapping regions of
spacetime.''. Correspondingly, this statement is inconsistent with the
explicit construction in the case of a magnetic monopole given in
Eqs.(10) and (11) of Ref.~\cite{Wu:1975es}, where it is essential that
$\delta > 0$. The limit $\delta \to 0$ cannot be taken.

In Refs.~\cite{Gonuguntla:2023cik,Gonuguntla:2023uxi} the authors attempt
to describe a magnetic monopole using various forms of a vector potential
defined on all of space (minus the origin). In all cases the resulting
magnetic field is not just the field of a magnetic monopole, but contains
additional contributions. We did not check the details of these calculations.
However, the fact that such additional contributions appear is in
accordance with Theorem 2 in Ref.~\cite{Wu:1975es} (after regularization of
the singularities (discontinuities) in the potentials).

In more mathematical terms, an electromagnetic field is described by a
connection on a U(1) principal fiber bundle, cf.\ Table I in
Ref.~\cite{Wu:1975es}. Trying to describe a magnetic monopole by a
single vector potential on all of space (minus the origin) means that
one works on a trivial fiber bundle. However, as explained in detail
in Ref.~\cite{Wu:1975es}, in the case of a magnetic monopole one needs
a nontrivial fiber bundle. The corresponding vector potentials are not
defined on all of space, but only on suitable open subsets of space
(or spacetime), see Eqs.(10) and (11) in Ref.~\cite{Wu:1975es}. On the
overlap of these regions the potentials are related by a gauge transformation,
whence the field strength is uniquely defined and given by that of a
magnetic monopole. The potentials are not glued together to
form a single potential on the whole space (the base space of the fiber
bundle). Rather they are local representatives of a global connection
(i.e.\ smoothly defined on the whole bundle space) on a nontrivial U(1)
principal bundle.

In summary: The calculations presented in
Refs.~\cite{Gonuguntla:2023cik,Gonuguntla:2023uxi} may be of some interest
in their own right, but they do not correspond to the
Wu-Yang fiber bundle approach~\cite{Wu:1975es}.

\end{document}